\documentclass{aa}  
\usepackage{graphicx}
\usepackage{txfonts}
\begin{document}
   \title{
	 An asteroseismic study of the $\delta$ Scuti star 44 Tau
   }

   \subtitle{}

   \author{
	 P. Lenz\inst{1}, 
	 A. A. Pamyatnykh\inst{1}\fnmsep\inst{2}\fnmsep\inst{3}, 
	 M. Breger\inst{1},
	 V. Antoci\inst{1}
   }

   \authorrunning{Lenz et al.}

   \offprints{P. Lenz}

   \institute{
	 Institut f\"ur Astronomie, University of Vienna,
     T\"urkenschanzstrasse 17, A-1180 Vienna, Austria\\
	 \email{lenz@astro.univie.ac.at}
     \and
     Copernicus Astronomical Center, Polish Academy of Sciences,
	 Bartycka 18, 00-716 Warsaw, Poland
     \and
     Institute of Astronomy, Russian Academy of Sciences, 
	 Pyatnitskaya Str. 48, 109017 Moscow, Russia
   }

   \date{Received September 15, 1996; accepted March 16, 1997}

 
  \abstract
   {}
   {In this paper we investigate theoretical pulsation models for the $\delta$ Scuti star 44~Tau. The star was monitored during several multisite campaigns which confirmed the presence of radial and nonradial oscillations. Moreover, its exceptionally low rotational velocity makes 44~Tau particulary interesting for an asteroseismic study. Due to the measured $\log g$ value of $3.6 \pm 0.1$, main sequence and post-main sequence models have to be considered.}
   {We perform mode identification based on photometric and spectroscopic data. A nonadiabatic pulsation code is used to compute models that fit the identified modes. The influence of different opacity tables and element mixtures on the results is tested.
   }
   {The observed frequencies of 44 Tau can be fitted in both the main sequence and the post-main sequence evolutionary stage. Post-main sequence models are preferable as they fulfill almost all observational constraints (fit of observed frequencies, position in the HRD and instability range). These models can be obtained with normal chemical composition which is in agreement with recent spectroscopic measurements. The efficiency of envelope convection (in the framework of the mixing-length theory) is predicted to be very low in 44 Tau. We show that the results are sensitive to the choice between the OPAL and OP opacities. While the pulsation models of 44 Tau computed with OP opacities are considerably too cool and too faint, the use of OPAL opacities results in models within the expected temperature and luminosity range.
   }
   {}

   \keywords{
	 stars: variables: $\delta$ Sct --
	 stars: oscillations --
	 stars: rotation --
	 stars: individual: 44~Tau
   }

   \maketitle


\section{Introduction}
\label{sec:intro}

$\delta$ Scuti stars are pulsating stars situated at the lower part of the classical instability strip in the HR diagram. Their spectral types range from A0 to F2 and they are in the hydrogen-core or hydrogen-shell burning stage. With masses between 1.5 and 2.5~M$_{\odot}$ their periods of pulsation are between 30 min and 8~hours. In these stars pulsation is driven by the opacity mechanism ($\kappa$ mechanism) acting in the HeII ionization zone.

Among the $\delta$ Scuti stars two subgroups are distinguished: high-amplitude $\delta$ Scuti stars (HADS) with peak-to-peak amplitudes $\geq$ 0.3 mag in $V$ with mainly radial pulsation and low-amplitude $\delta$ Scuti stars which oscillate predominantly non-radially. HADS stars show rather slow rotation in comparison with the average $v \sin i$ of $96~\rm km~s^{-1}$ of low amplitude $\delta$ Scuti stars (Solano \& Fernley \cite{solano}). 44~Tau (HD 26322, spectral type: F2 IV) exhibits characteristics of both groups: radial as well as nonradial modes are excited to observable amplitudes and its projected rotational velocity is very small ($v \sin i = 2 \pm 1~\rm km~s^{-1}$).  As discussed by Antoci et al. (\cite{antoci}) (hereafter ABR07) this indicates either pole-on view and/or intrinsic slow rotation. Zima et al. (\cite{zima}) recently confirmed that 44~Tau is an intrinsic slow rotator. This makes this star particulary interesting for asteroseismology.

Previous theoretical studies on this star have been carried out by Civelek et al. (\cite{civelek}) and K{\i}rb{\i}y{\i}k et al. (\cite{kirbiyik}). By means of fitting adiabatic oscillation frequencies to the observed data they conclude that 44~Tau is a 1.89 M$_{\odot}$ star at the hydrogen-shell burning stage. Recently, Garrido et al. (\cite{garrido}) concluded that the pulsations of 44~Tau could be explained by a main sequence model with a mass of 1.94 M$_{\odot}$. 

This paper extends the work of Lenz et al. (\cite{lenz}) and presents a comprehensive seismic study based on the detailed frequency analysis of ABR07 and on additional photometric data obtained in 2004/05.

The paper is structured into the following parts: Section~\ref{sec:obs} summarizes the results of available photometric and spectroscopic data. Section~\ref{sec:models} is devoted to the theoretical modelling of the frequencies of 44~Tau. After the determination of the spherical degrees of the observed frequencies we compute models to fit the detected frequencies and the observed frequency range of excited modes. Finally, the effect of using different opacity tables and element mixtures is studied. In Section~\ref{sec:conclusions} we present our conclusions.


\section{Observations}
\label{sec:obs}

\subsection{Photometry}

\begin{table*}
  \caption{
Averaged observed amplitude ratios and phase differences from 2000/1 to 2004/5. Only frequencies detected in both photometric and spectroscopic data are listed. In the last two columns the predicted spherical degree from our mode identification (see Section~\ref{sec:modeID}) and the azimuthal number $m$ from Zima et al. (2007) are given. For f$_9$ no reliable results could be derived due to the influence of the close and more dominant frequency f$_6$.}
  \label{table:1}
  \centering
  \begin{tabular}{lrrrrr}
	\hline
	 & Frequency     & A$_v$/A$_y$  & $\phi_v$-$\phi_y$  & $\ell$ & $m$ \\
	 & [cd$^{-1}$]  & [mmag]       &  [$^\circ$]        &     &        \\
	\hline
	f$_1$ & 6.8980& 1.44 $\pm$ 0.01 & +2.96 $\pm$ 0.22 & 0& 0\\ 
	f$_2$ & 7.0060& 1.43 $\pm$ 0.01 & -1.83 $\pm$ 0.39 & 1& 1\\ 
	f$_3$ & 9.1175& 1.46 $\pm$ 0.02 & -1.85 $\pm$ 0.29 & 1& 1\\ 
	f$_4$ &11.5196& 1.42 $\pm$ 0.01 & -1.75 $\pm$ 0.50 & 1& 0\\
	f$_5$ & 8.9606& 1.45 $\pm$ 0.03 & +2.35 $\pm$ 0.36 & 0& 0\\
	f$_6$ & 9.5613& 1.46 $\pm$ 0.02 & -1.40 $\pm$ 0.32 & 1&  \\
	f$_7$ & 7.3034& 1.45 $\pm$ 0.06 & -6.75 $\pm$ 1.61 & 2& 0\\ 
	f$_8$ & 6.7953& 1.38 $\pm$ 0.08 & -7.26 $\pm$ 4.57 & 2& 0\\
	f$_9$ & 9.5801& -               & -                &  &  \\
	f$_{10}$& 6.3391& 1.40 $\pm$ 0.10 & -4.97 $\pm$ 5.69 & 2& \\
	f$_{11}$& 8.6393& 1.35 $\pm$ 0.09 & -7.44 $\pm$ 1.58 &  & 0\\
	f$_{12}$&11.2919& 1.32 $\pm$ 0.19 & -3.21 $\pm$ 5.35 &  &  \\ 
	\hline
  \end{tabular}
\end{table*}

44~Tau was the main target of several Delta Scuti Network (DSN) campaigns. Photometric data obtained during three campaigns between 2000 and 2003 were analysed in detail by ABR07. 29 significant frequencies of which 13 are independent were detected. 

In the winter season 2004/05 an additional observing run was dedicated to this star. The new photometric data confirmed the previous results. Although the noise for the total data set is much lower only a few additional frequencies could be detected. However, these frequencies are combinations of already detected frequencies and no additional independent frequency was found.

\begin{figure}[htb]
  \centering
  \includegraphics[width=8cm]{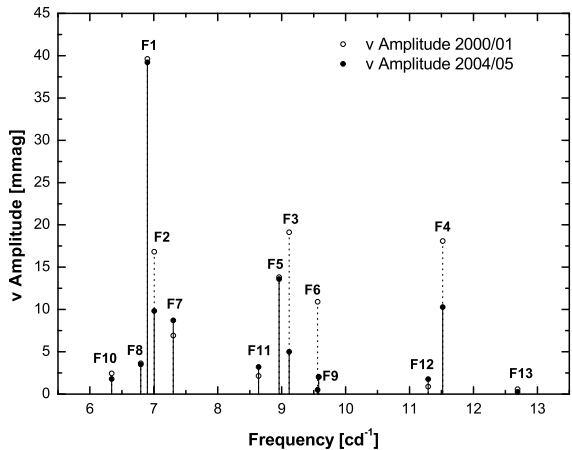}
  \caption{Comparison of Str\"omgren $v$ amplitudes from the observing season 2000/01 with the data from 2004/05. The error bars are smaller than the size of the given points. 
 The frequencies f$_2$, f$_3$, f$_4$ and f$_6$ show strong amplitude variation.}
  \label{fig:photometry}
\end{figure}

Some modes in 44~Tau show strong amplitude variability. This is not unusual and is a rather common phenomenon in $\delta$~Scuti stars (Poretti \cite{poretti2000}, Breger \cite{ampVar4CVn}).
 There are several possible explanations such as, for example, beating of close frequencies or resonance effects. In Fig.~\ref{fig:photometry} the frequency spectrum of the 2004/05 data is compared to the results from the 2000/01 data set. The presence of amplitude variations is clearly visible.

\subsection{Spectroscopy}

In 2004 a multisite spectroscopic campaign was carried out. Zima et al. (\cite{zima}) detected 16 frequencies of which 4 are combinations. 12 independent frequencies were found in the spectroscopic as well as in the photometric data. Using the moment method they constrained the azimuthal order of 8 modes. f$_1$, f$_4$, f$_5$, f$_7$, f$_8$ and f$_{11}$ were identified as axisymmetric modes whereas f$_2$ and f$_3$ were found to be prograde modes with $m=1$. The corresponding frequency values are listed in Table~\ref{table:1}.

Only two modes were determined to be non-axisymmetric, therefore only a rough determination of the inclination angle $i$ was possible. Zima et al. (\cite{zima}) estimated $i$ to range between $35^\circ$ and $85^\circ$. Since $v \sin i$ is derived to be 2 $\pm$ 1 km~s$^{-1}$ this would correspond to an equatorial rotational velocity  between 1 and 5 km~s$^{-1}$. Note that the newly derived value for the projected rotational velocity is slightly lower than the results of previous studies by Solano \& Fernley (\cite{solano}) who measured $v \sin i$ to be $6.8 \pm 1.2$ km~s$^{-1}$.

The analysis of the photospheric abundances of 44~Tau by Zima et al. (\cite{zima}) shows a similar composition as for the sun. The effective temperature is estimated to be 7000 $\pm$ 200 K and $\log g$ is determined to be $3.6 \pm 0.1$ consistently from line profile and equivalent width measurements.


\section{Seismic Modelling}
\label{sec:models}

\subsection{Fundamental Parameters of 44~Tau}

Based on the measured HIPPARCOS parallax of $16.72~\pm~0.93~\mbox{mas}$, a luminosity of $\log L/L_{\odot}$~=~1.340~$\pm$~0.065 was derived.  The effective temperature was estimated from Str\"omgren and Geneva photometry and from the Vienna New Model Grid of Stellar Atmospheres (NEMO, Nendwich et al. \cite{nendwich}, Heiter et al. \cite{heiter}). The derived value is $\log T_{\rm eff}$~=~3.839~$\pm$~0.007 (6900~K~$\pm$~100~K). This result is in agreement with the value inferred from spectroscopy.

The spectroscopically measured $\log g$ value of $3.6 \pm 0.1$ does not allow  an unambiguous determination of the evolutionary status of 44~Tau. This can be clearly seen in a $\log g$ vs. $\log T_{\rm eff}$ plot such as Fig. 5 in Pamyatnykh (\cite{pam2000}). Consequently, both main sequence and post-main sequence models  have to be considered.

Since the intrinsic rotational velocity of 44 Tau is very low,
we consider only nonrotating models in our study.

\subsection{Tools}
\label{sec:tools}

The models in this paper were computed with an improved version of the codes described by Breger \& Pamyatnykh (\cite{bregerpam}). The OPAL opacity tables (Iglesias \& Rogers, \cite{iglesias}) supplemented with the low-temperature data of Alexander \& Ferguson (\cite{alexander}) were used. If not stated otherwise, we made use of the element mixture from Grevesse \& Noels (\cite{gn93}, hereafter GN93) and the last version of the OPAL equation of state (see Rogers et al. \cite{rogers}, version OPAL EOS 2005). Nuclear reaction rates are taken from Bahcall and Pinsonneault (\cite{bahcall}). A linear nonadiabatic analysis of low-degree oscillations was performed using a modified version of Dziembowski's code (1977).

\subsection{Radial Modes}

In ABR07 the radial character of the frequencies f$_1$ and f$_5$ was confirmed. The fact that radial fundamental and first overtone are observed can be used to reduce the number of possible models significantly. To find appropriate models for 44~Tau  it is useful to utilize Petersen diagrams (Petersen \& Jorgensen \cite{petersen}). In the Petersen diagram the period ratio of the radial first overtone to the fundamental mode is plotted as a function of the period of the fundamental mode. These diagrams can be used to determine the correct mass of the model which fits the radial fundamental frequency and the ratio F/1H (f$_1$/f$_5$=0.7698) to the observed values. The period ratio is sensitive to metallicity and rotation as shown by Su\'arez et al. (\cite{suarez}). In Section~\ref{sec:opac} we will demonstrate the effect of different opacity tables and the new solar element mixture on the period ratio.

\begin{figure}
  \centering
  \includegraphics[width=7.7cm]{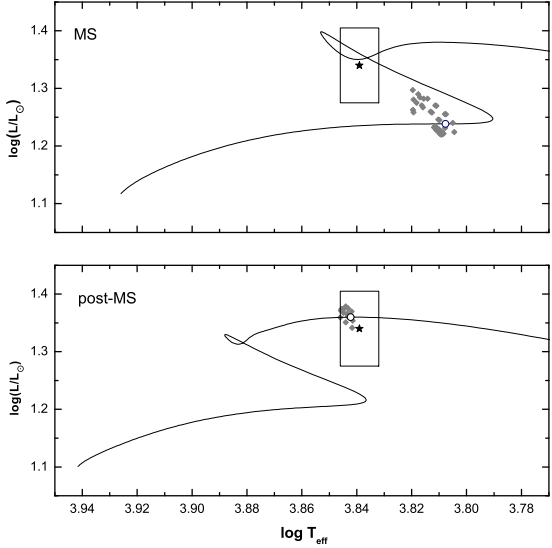}
  \caption{The Hertzsprung-Russell diagram showing the location of models that fit radial fundamental and first overtone modes to the observed frequencies for the main sequence (upper panel) and post-main sequence case (lower panel). The asterisk indicates the center of the photometric error box. Grey points represent models with Z values between 0.019-0.04 (MS) or 0.015-0.025 (post-MS) and an overshooting parameter $\alpha_{\rm ov}$ ranging from 0 to 0.4. Representative evolutionary tracks for the models listed in Table~\ref{table:bestmodels} are plotted.}
  \label{fig:hrd_combined}
\end{figure}

\begin{figure}
  \centering
\includegraphics[width=8cm]{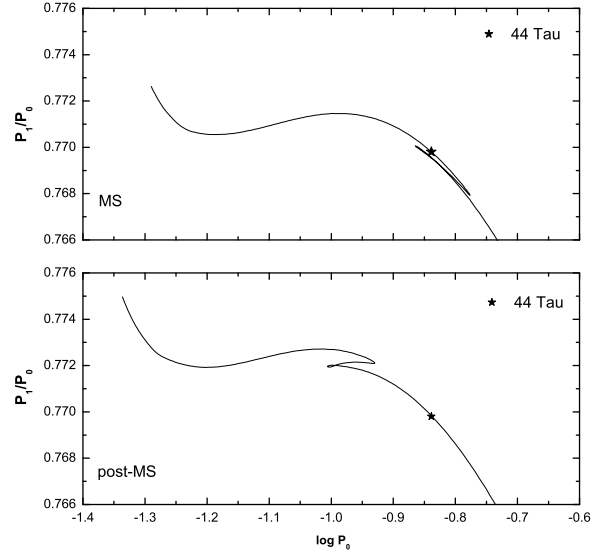}
  \caption{Petersen diagrams for a main sequence (upper panel) and a post-main sequence model (lower panel). The asterisk corresponds to the observed values of 44~Tau. The uncertainties in period ($\sim 1.4 \times 10^{-5}$ cd$^{-1}$) and period ratio ($\sim 2 \times 10^{-6}$) are smaller than the symbol size.
  }
  \label{fig:petersen_combined}
\end{figure}

We have computed stellar models for a large set of different input parameters such as chemical composition, mixing-length parameter, $\alpha_{\rm MLT}$, and the parameter for overshooting from the convective core, $\alpha_{\rm ov}$. Our results show that main sequence and post-main sequence models which fit the radial fundamental and first overtone modes are clearly separated in the HR diagram (see Fig.~\ref{fig:hrd_combined}). For each case an evolutionary track of a selected model is delineated. Post-main sequence models are located inside the photometric error box, whereas main sequence models are predicted at significantly lower effective temperatures in the HR diagram. 
The given family of models was computed using a wide range of parameters, such as Z between 0.015 and 0.04 and with an overshooting parameter $\alpha_{\rm ov}$ of up to 0.4 pressure scale heights. As a representative example, the Petersen diagram for two selected models is shown in Fig.~\ref{fig:petersen_combined}. The corresponding model parameters are listed in Table~\ref{table:bestmodels}. 

\begin{table*}
  \caption{Physical parameters of two representative models used in this paper. If no indication is given in figures the corresponding plots refer to the models in this table.}
  \label{table:bestmodels}
  \centering
  \begin{tabular}{llllllllll}
	      & X & Z & $\alpha_{\rm MLT}$ & $\alpha_{\rm ov}$ & M/M$_{\odot}$ & $\log T_{\rm eff}$ & $\log L$ & $\log g$ & Age [Myr] \\
	\hline
	MS     & 0.70 & 0.03 & 0.0 & 0.25 & 2.010 & 3.8077 & 1.2385 & 3.6848 & 1150\\
	post-MS& 0.70 & 0.02 & 0.0 & 0.0  & 1.875 & 3.8422 & 1.3601 & 3.6712 & 1120\\
	\hline
  \end{tabular}
\end{table*}

\subsection{Identification of Nonradial Modes}
\label{sec:modeID}

In the previous section we have noted that the observation of two radial modes puts strong constraints on the models. For complete seismic modelling it is necessary to obtain an identification for a sufficient number of observed nonradial pulsation modes. 

Mode identification is often complicated by the effects of rotation. The exceptionally small rotational velocity of 44~Tau allows to reduce these problems because the rotational splitting is very small. There are several techniques that either make use of photometric data in two filters or combine them with spectroscopic data to infer the spherical degrees of the modes. The latter approach has been successfully applied in the case of FG~Vir (Daszy\'{n}ska-Daszkiewicz et al. \cite{jdd2005}). The azimuthal order of pulsation modes can only be determined from spectroscopy. 

To obtain reliable photometric mode identification we computed average amplitude ratios and phase differences in the Str\"omgren filters $v$ and $y$ using weights for each observing season to account for the different amount and quality of the data (see formulae in the Appendix of Breger et al. \cite{breger1999}). The observational uncertainties were derived by error propagation from the uncertainties of the annual solutions. These results are given in Table~\ref{table:1}.

\begin{figure}
  \centering
  \includegraphics[width=8cm]{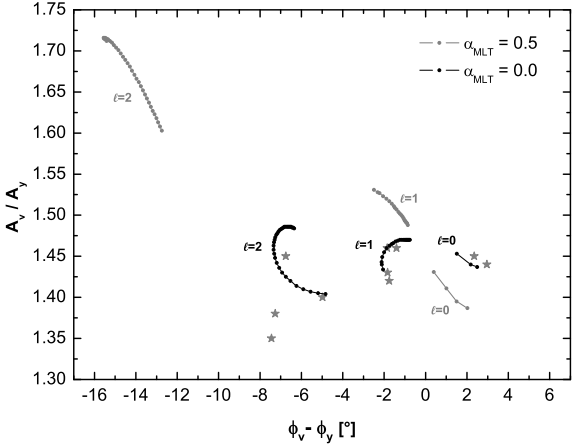}
  \caption{Diagnostic diagram used for mode identification. Comparison of theoretical results for $\alpha_{\rm MLT}=0.5$ and  $\alpha_{\rm MLT}=0.0$ to the observational positions.  The asterisks correspond to the amplitude ratios and phase differences of the observed frequencies. Theoretical mode positions are given as points connected with lines. Only predicted modes within the observed frequency range are plotted. }
  \label{fig:modeIDalpha}
\end{figure}

\begin{figure*}
  \centering
  \includegraphics[width=8cm]{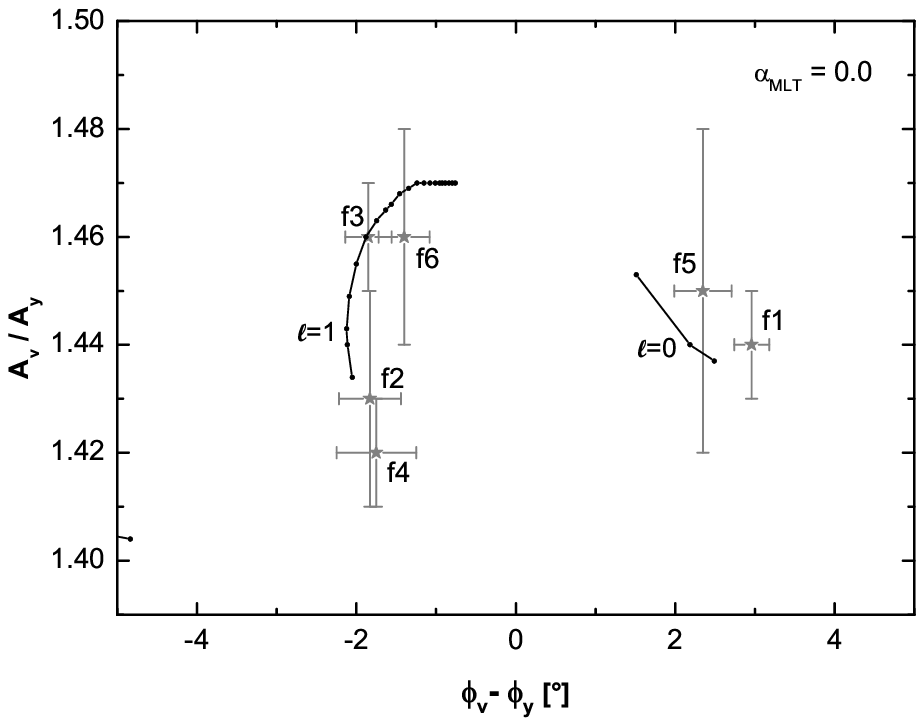}
  \hspace{1.5cm}
  \includegraphics[width=8cm]{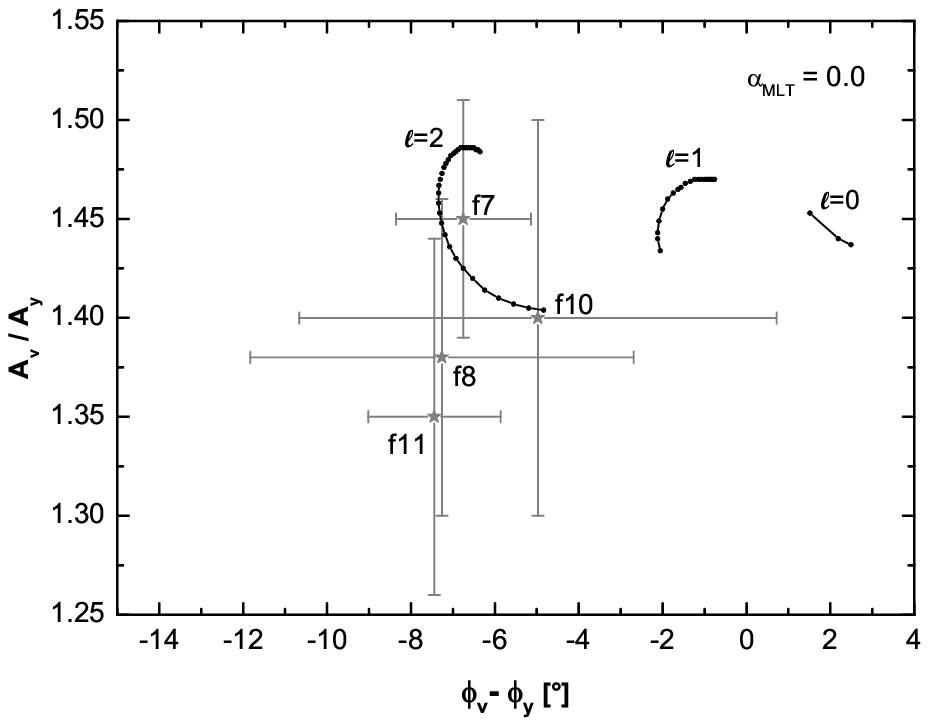}
  \caption{Observed mode positions with error bars. For better clarity modes with small and large error bars are shown in separate diagrams. In both diagrams the predicted mode positions (for modes within the observed frequency range) refer to the case of inefficient convection, $\alpha_{\rm MLT}=0.0$. The diagrams plotted for a stellar model with 1.85 M$_{\odot}$, X=0.70, Z=0.02 and $\alpha_{\rm ov}=0.0$.}
  \label{fig:modeID}
\end{figure*}

For $\delta$ Scuti stars theoretical mode positions in the amplitude ratio vs. phase difference diagram are sensitive to the efficiency of convection in the stellar envelope. Montalb\'{a}n \& Dupret (\cite{montalban}) showed that different treatments of convection affect the phase differences and amplitude ratios as well. 
In our computations, we use the mixing-length theory of convection and assume the frozen convective flux approximation for the pulsation-convection interaction. This assumption may not be adequate for cold $\delta$ Scuti stars and result in large uncertainties in the mode identification. However, Montalb\'{a}n \& Dupret (\cite{montalban}) find that for frequencies close to that of the radial fundamental mode the convection-pulsation treatment does not influence the determination of mode degree, $\ell$.

We tested different values of the mixing-length parameter $\alpha_{\rm MLT}$ for a stellar model with standard chemical composition and a mass of 1.85 M$_{\odot}$. The results of these computations are shown in Fig.~\ref{fig:modeIDalpha} and~\ref{fig:modeID} . The theoretical mode positions were calculated according to Daszy\'{n}ska-Daszkiewicz et al. (\cite{jdd2003}) using photometric data in two passbands.

As can be seen in Fig.~\ref{fig:modeIDalpha}, for $\alpha_{\rm MLT}=0.5$ (or higher) there is no reliable agreement between observed and theoretical mode positions. Instead, we obtain a good consistency for inefficient convection ($\alpha_{\rm MLT} \lesssim 0.2$).
 In the two diagrams in Fig.~\ref{fig:modeID} the positions of observed modes are compared to the theoretical predictions for a model with inefficient convection. We can definitively identify the frequencies f$_1$ and f$_5$ as $\ell$=0 modes, f$_2$, f$_3$, f$_4$ and f$_6$ as $\ell$=1 modes. f$_7$ can be identified as an $\ell=2$ mode. The observational formal errors for the remaining frequencies are too large to draw a conclusion. However, an identification as $\ell=2$ is probable for some of the modes. Still, it is necessary to confirm their mode degree by another method.

Since radial velocity data from 2004 are available (Zima et al. \cite{zima}), it is possible to combine them with photometric data from the Str\"omgren $v$ and $y$ passbands using the method by Daszy\'{n}ska-Daszkiewicz et al. (\cite{jdd2003}, \cite{jdd2005}). This method allows to perform a $\chi^2$ test which indicates the most probable spherical degree for a mode. The results 
agree with those from the purely photometric mode identification. Additionally this method identifies f$_{8}$ and f$_{10}$ as $\ell=2$ modes with a high confidence level.
Table~\ref{table:1} summarizes the results of our mode identification.

As mentioned before, we rely on the assumption of frozen convective flux, which may not be a realistic assumption for cold $\delta$ Scuti stars. A comparison with mode identification results using a time-dependent convection approach is thus desirable. Recently Garrido et al. (\cite{garrido}) computed diagnostic diagrams to identify the pulsation modes of 44~Tau using the time-dependent convection approach described by Dupret et al. (\cite{dupret2004}) and Grigahcene et al. (\cite{griga2005}). Even though  using different treatments of convection their results are in agreement with our results. 
Within the same context but for another $\delta$ Scuti star (FG Vir) Daszy\'{n}ska-Daszkiewicz et al. (\cite{jdd2005}) have compared the results for the complex parameter $f$ and mode instability from the frozen convective flux assumption and the nonlocal time-dependent mixing-length theory (Gough \cite{gough1977a}). As for 44~Tau, envelope convection is inefficient in FG Vir and the results from both treatments are similar. Consequently, our simple approach with a small mixing-length parameter ($\alpha_{\rm MLT} \lesssim 0.2$) may also provide reliable results in the case of 44~Tau.

By comparing our mode identification results with the frequency spectrum in Fig.~\ref{fig:photometry}, an interesting conclusion can be drawn: the strongest amplitude variability is shown by $\ell=1$ modes, whereas the two radial modes are constant in their amplitudes. This fact may be important to explain the cause for the observed amplitude variability in 44~Tau.

\begin{figure*}[htb]
  \centering
  \includegraphics[width=16cm]{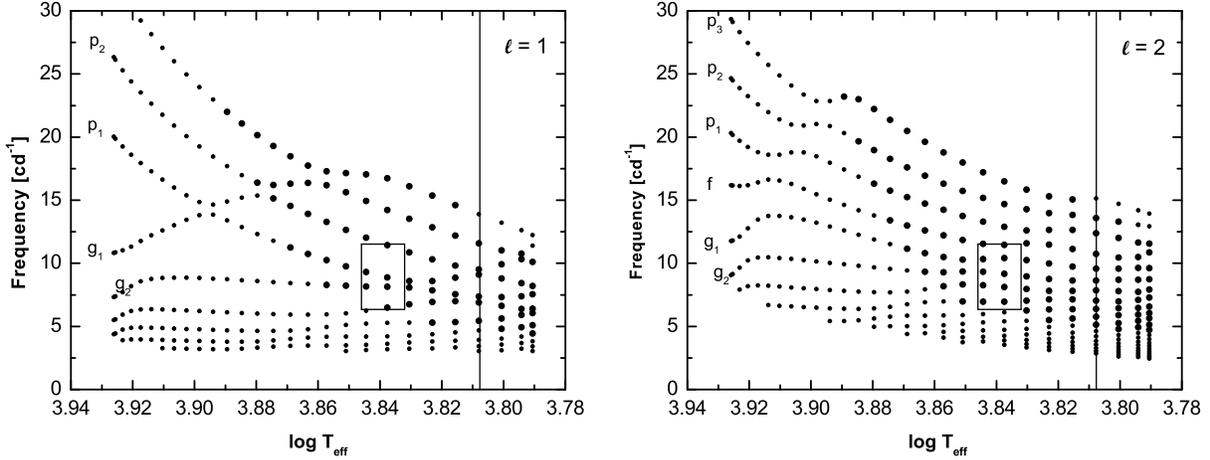}
  \caption{Behaviour of the frequencies of $\ell=1$ modes (left panel) and $\ell=2$ modes (right panel) in the course of the main sequence evolution from the ZAMS (leftmost points) to the TAMS (rightmost points). Larger points represent unstable modes. The vertical line represents the effective temperature of the model with the best fit to the observed frequencies. The box indicates the ranges for observed frequencies and for effective temperature determined from photometry.
 The parameters of the model are given in Table~\ref{table:bestmodels}.}
  \label{fig:freqevolution}
\end{figure*}

Two of the $\ell$=1 modes, f$_3$ and f$_6$, are separated by only 0.44 cd$^{-1}$ whereas the frequency distance between the other $\ell$=1 modes f$_2$ and f$_3$ or f$_6$ and f$_4$ is around 2 cd$^{-1}$. This pattern indicates that we observe an avoided crossing phenomenon (Aizenman et al.~\cite{aizenman}), which is a coupling between two modes of different type, viz. an acoustic (p) and a gravity (g) mode. If an avoided crossing is indeed observed it represents an additional strong constraint to the model (see Dziembowski \& Pamyatnykh \cite{wadaap1991}).
In Fig.~\ref{fig:freqevolution} the evolutionary change of the frequencies is shown. The given box restricts the uncertainties of the measured effective temperature and the observed frequency range. Within this box an avoided crossing between $g_1$ and $g_2$ takes place (the $g_1$ mode behaves here like an acoustic $p$-mode due to the previous avoided crossing with the $p_1$ mode). For the main sequence model which fits the radial modes (vertical line) we predict two avoided crossing pairs for $\ell=1$ within the observed frequency range. One of them fits the observed frequencies as we will see later.

\subsection{Instability Ranges}

Successful seismic modelling also involves the correct prediction of instability of the detected modes. Instability reflects the properties of the envelope, where mode driving occurs. 

\begin{figure}[htb]
  \centering
  \includegraphics[width=9cm]{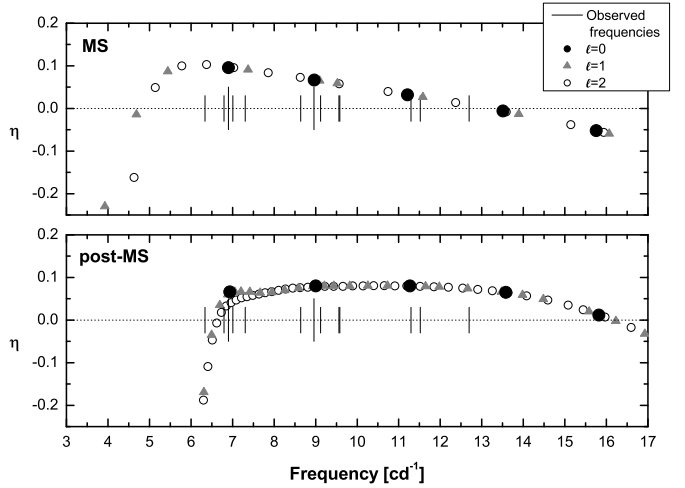}
  \caption{Comparison of the observed and theoretical frequency ranges for main sequence (upper panel) and post-main sequence (lower panel) models. The normalized growth-rate, $\eta$, is positive for unstable modes. The vertical lines represent the observed frequencies (with longer lines for $\ell=0$ modes).}
  \label{fig:instability}
\end{figure}

Instability ranges for two representative models (on the main sequence and 
in the post-main sequence stage) which fit the radial as well as $\ell$=1 modes are shown in Fig.~\ref{fig:instability}. In this diagram the normalized growth-rate as defined by Stellingwerf (\cite{stellwf:1978}), $\eta~=~\int_{\rm 0}^{\rm R}~({\rm d}W/{\rm d}r)~{\rm d}r~/~\int_{\rm 0}^{\rm R}~|{\rm d}W/{\rm d}r|~{\rm d}r$ (where W is the usual work integral), is plotted against frequency. $\eta$ varies between -1 (full damping) and +1 (full driving). It can be seen that for the main sequence model the instability range ($\eta>0$) covers the observed frequency range completely but extends to considerably lower frequencies than observed. For the post-main sequence model the instability range extends to higher frequencies than actually observed.
The theoretical lower boundary for unstable frequencies fits the observational limit much better than in the main sequence case but the lowest observed frequency f$_{10}$ is predicted to be stable.
 With exception of a significant effect of convective efficiency on the high frequency boundary of the instability range, the results are not very sensitive to changes of the input parameters. 

Generally, the value of the highest unstable frequency is very sensitive to the efficiency of convection and to the convection-pulsation treatment. Since our code relies on the simplest frozen convection assumption, it is not suitable to predict the upper limit of the frequency instability range.
If future observations reveal additional independent frequencies that extend the observed frequency range, further constraints on the evolutionary status and on the efficiency of convection in 44~Tau may be provided.


\subsection{Main Sequence Models}

As discussed in Section~\ref{sec:tools}, it is not possible to fit the observed period ratio P$_1$/P$_0$ and the measured fundamental parameters simultaneously for models in the hydrogen-core burning stage. The predicted main sequence models for 44~Tau are generally too cool (see Fig.~\ref{fig:hrd_combined}). 
However, the photometric calibrations may have larger uncertainties than adopted in Fig.~\ref{fig:hrd_combined}. Therefore, we also consider MS models for 44 Tau. We computed a grid of models by varying the chemical composition and the overshooting efficiency. Not for all models the radial fundamental and first overtone mode could be fitted on the main sequence.
Using standard chemical composition (X=0.70, Z=0.02) and assuming inefficient convection, $\alpha_{\rm MLT}=0.0$, it is not possible to obtain any main sequence model for $\alpha_{\rm ov} < 0.25$. For $\alpha_{\rm ov} = 0.25$ the model is located exactly at the TAMS. 
For Z=0.019 and 0.021 main sequence models are possible for $\alpha_{\rm ov} \geq$ 0.28 and 0.23, respectively. Thus, the presence of effective convective overshooting is essential to model 44~Tau in the stage of hydrogen burning in the core.

\begin{figure}
  \centering
  \includegraphics[width=9cm]{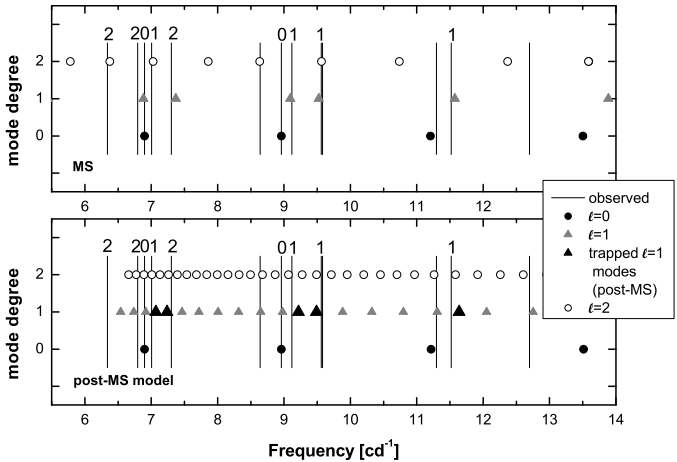}
  \caption{Comparison between observed and predicted frequencies for a main sequence (upper panel) and post-main sequence (lower panel) case. Observed frequencies are marked by vertical lines. The numbers above the lines denote the corresponding mode degree, $\ell$, as identified from photometry and spectroscopy.}
  \label{fig:freqs}
\end{figure}

The evolutionary track for the main sequence model with the best fit of the identified observed frequencies is delineated in the upper panel of Fig.~\ref{fig:hrd_combined}. This 2.01 $M_{\odot}$ model was computed using X=0.70, Z=0.03, $\alpha_{\rm MLT}$=0.0, $\alpha_{\rm ov}$=0.25. Additional information on the physical parameters of the model is given in Table~\ref{table:bestmodels}. The predicted frequency spectrum (see Fig.~\ref{fig:freqs}, upper panel) shows good agreement between predicted and observed frequencies of $\ell=0$ and $\ell=1$ modes. A good fit of the $\ell=2$ modes could not be obtained.

\subsection{Post-Main Sequence Models}

In contrast to main sequence models it is not difficult to obtain models within the photometric error box in the post-main sequence case (Fig.~\ref{fig:hrd_combined}, lower panel). The post-main sequence evolution across the observed error box is only 5-9 times faster than the main sequence evolution. Therefore, the probability to observe a star in the H-shell burning stage is not small.
 For the representative model given in Fig.~\ref{fig:hrd_combined}, the input parameters X=0.70, Z=0.02, $\alpha_{\rm MLT}$=0.0, $\alpha_{\rm ov}$=0.0 and an initial mass of 1.875 M$_{\odot}$ were chosen. The corresponding frequency spectrum for this model which fits the observed radial modes can be seen in Fig.~\ref{fig:freqs} (lower panel).

\begin{figure}[htb]
  \centering
  \includegraphics[width=8cm]{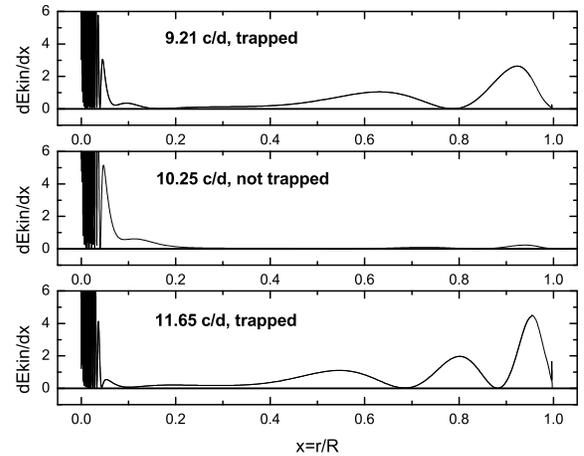}
  \caption{Kinetic energy density of pulsation along the stellar radius for three $\ell=1$ modes. The modes at 9.21 and 11.65 cd$^{-1}$ are trapped, whereas the mode at 10.25 cd$^{-1}$ is not. The energy contribution from the outer envelope is significant for the two trapped modes.}
  \label{fig:trappeigenfunc}
\end{figure}

The frequency spectrum of nonradial modes is much denser than that of main sequence models. In fact, for post-main sequence models many more frequencies are predicted than observed. An extreme example is 4 CVn, one of the most evolved $\delta$ Scuti stars, for which approximately 500 modes are predicted (Breger \& Pamyatnykh \cite{breger4CVn}).
There certainly exist some rules which select only a few modes to be excited to observable amplitudes. Mode trapping in the acoustic cavity in the stellar envelope can be one of these rules, as discussed by Dziembowski \& Krolikowska (\cite{wadkro1990}).

Trapped modes can be considered as nonradial counterparts of radial modes. Most of the energy of trapped modes is confined in the envelope, where the density is low. Therefore, the total kinetic energy of these modes is much smaller than that of modes which are not trapped. Consequently, it may be easier to excite trapped modes to observable amplitudes. 
In Fig.~\ref{fig:trappeigenfunc} the distribution of the kinetic energy density of pulsation along the stellar radius for two trapped $\ell$=1 modes and for a mode which is not trapped is shown.

\begin{figure}[htb]
  \centering
  \includegraphics[width=8cm]{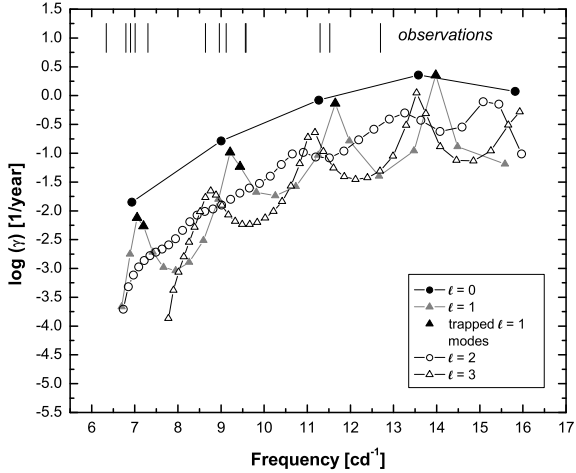}
  \caption{Increment of the exponential amplitude growth, $\gamma$, for unstable modes of $\ell \leq 3$ for the post-MS model shown in Fig. 6. The trapped $\ell$=1 modes are marked with black triangles.}
  \label{fig:trapping}
\end{figure}

Mode trapping can also explain the tendency of detected frequencies to build groups around radial frequencies as can be seen in Fig.~\ref{fig:trapping}.
This diagram shows the increment of the exponential amplitude growth, $\gamma=-\mbox{Im}(\omega)$, where $\omega$ is the complex eigenfrequency. $\gamma$ is proportional to the work integral and inversely proportional to the kinetic energy of a given mode. Just the kinetic energy is responsible for its nonmonotonic behavior. The increment $\gamma$ is significantly higher for trapped modes. 

\begin{figure}[htb]
  \centering
  \includegraphics[width=8cm]{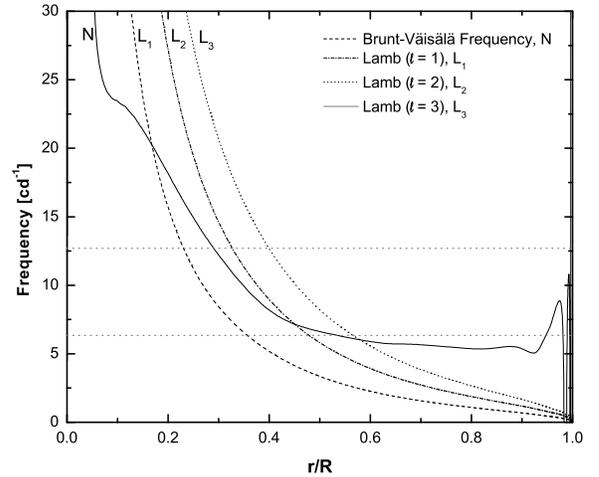}
  \caption{Run of the Brunt-V\"ais\"al\"a and Lamb-frequencies for $\ell \leq 3$ inside the post-MS model of 44 Tau. The two horizontal lines mark the observed frequency range ($\nu$ between 6.3 and 12.7 cd$^{-1}$). The extent of the evanescent zone is small for $\ell=2$ resulting in only weak trapping of $\ell=2$ modes. }
  \label{fig:bvplot}
\end{figure}

We can also see, that trapping of $\ell=2$ modes is rather weak compared to the $\ell=1$ and $\ell=3$ case (see Fig.~\ref{fig:trapping}). The reason for this becomes clear when the run of the Brunt-V\"ais\"al\"a frequency, $N$, and the Lamb frequency, $L$, is examined in detail. These characteristic frequencies determine the regions with the oscillatory behaviour in the star as shown in Fig.~\ref{fig:bvplot}. Similar diagrams can be found, for example, in Unno et al. 1979 (see Fig. 15.1 there). The propagation zone for $p$-modes is defined by the conditions $\nu^2 > N^2$ and $\nu^2 > L^2$ where $\nu$ ($=\mbox{Re}(\omega)/(2\pi)$~) denotes the mode frequency. $g$-modes are confined mainly in stellar regions where $\nu^2 < N^2$ and $\nu^2 < L^2$. Between these two cavities there is a so-called evanescent zone with exponentially decreasing amplitude behaviour.
While the Brunt-V\"ais\"al\"a frequency is independent of the spherical degree, the Lamb frequency incorporates the factor $\sqrt{\ell (\ell+1)}$. As can be seen in Fig.~\ref{fig:bvplot}, the mode cavities of $p$- and $g$-modes are poorly separated for $\ell=2$ modes in the observed frequency range (note the very small zone between N and L$_2$, in contrast to the more extended zones between L$_1$ and N for $\ell=1$ and between N and L$_3$ for $\ell=3$). Therefore, no effective mode trapping occurs for $\ell=2$ modes. 

This might explain why $\ell=2$ modes are observed with relatively small amplitudes only in 44 Tau as well as in other $\delta$ Scuti stars (such as FG Vir and 4 CVn).

\subsection{Effect of Opacities and Element Mixture}
\label{sec:opac}

As rotational effects can be neglected due to the low measured rotational velocity of about 3 $\pm$ 2 km s$^{-1}$, 44 Tau is a good target to test the effects of different input data on the Petersen diagram.

The results in the preceding sections were obtained with the OPAL opacity tables. Another, completely independent set of opacity tables is available from the Opacity Project (OP, see Seaton \cite{seaton:2005}). 
The use of different opacity tables slightly influences the location and size of the opacity bumps inside stellar models and therefore influences the driving of pulsations. Due to differences in opacities we expect also differences in the stellar structure and in the predicted frequencies of pulsations.

\begin{figure}[htb]
  \centering
  \includegraphics[width=8cm]{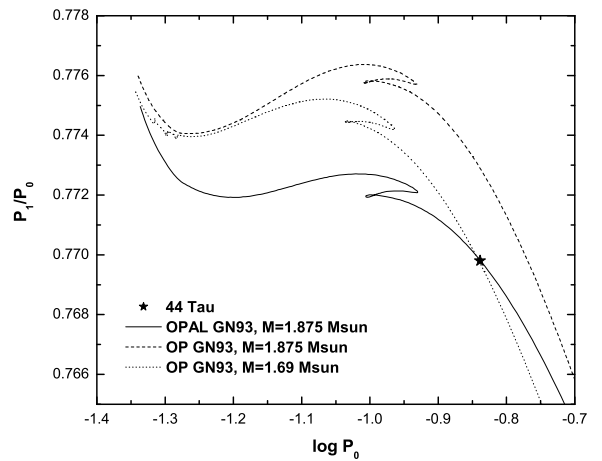}
  \caption{Comparison of the results for OPAL and OP opacities in the Petersen diagram in the case of a nonrotating post-main sequence model with standard chemical composition, inefficient convection ($\alpha_{\rm MLT}=0.2$) and assuming no overshooting from the convective core. The mass for the OP model was reduced to fit the observed period ratio of 44~Tau. 
}
  \label{fig:petersen_opac}
\end{figure}

\begin{figure}[htb]
  \centering
  \includegraphics[width=8cm]{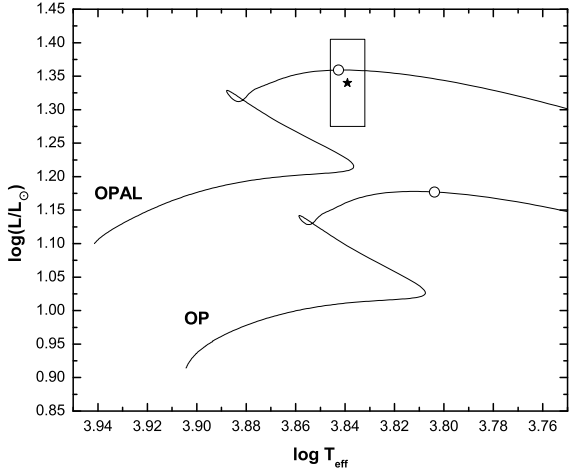}
  \caption{Comparison of the evolutionary tracks of the models given in Fig.~\ref{fig:petersen_opac}.}
  \label{fig:hrd_opac}
\end{figure}

We compared the preceding results with those obtained by using the OP tables. Using the same model parameters and element mixture (Grevesse \& Noels et al. \cite{gn93}) as for the standard OPAL post-main sequence model, we computed evolutionary tracks and corresponding frequencies. In the Petersen diagram (Fig.~\ref{fig:petersen_opac}) the difference of the period ratio between models computed with the two opacity sets is shown. The period ratios for the OP case are higher than those obtained with OPAL tables. Hence for models computed with OP opacities a lower mass is predicted for 44~Tau. However, as can be seen in Fig.~\ref{fig:hrd_opac}, seismic models computed with the OP opacities are clearly too cool and too faint. We do not consider main sequence models here, because their effective temperatures are already very low using OPAL tables (Fig.~\ref{fig:hrd_combined}, upper panel).

\begin{figure}[htb]
  \centering
  \includegraphics[width=8cm]{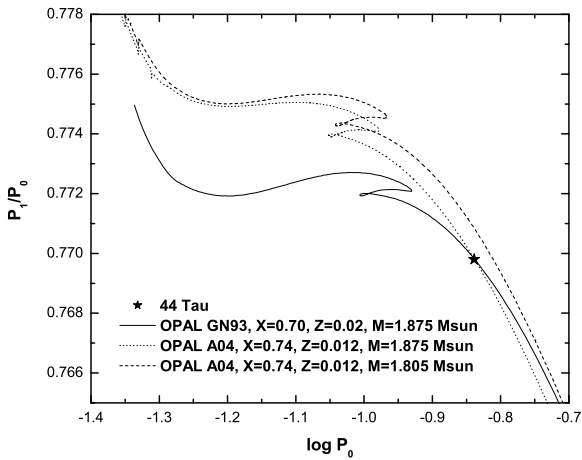}
  \caption{Comparison of period ratios for GN93 and A04 element mixtures for post-main sequence models obtained with OPAL opacities. For the models computed with A04 the new solar values X=0.74, Z=0.012 were used. Just as in Fig.~\ref{fig:petersen_opac}, a nonrotating post-main sequence model with inefficient convection ($\alpha_{\rm MLT}=0.2$) and no overshooting from the convective core was assumed.}
  \label{fig:pet_mix}
\end{figure}

\begin{figure}[htb]
  \centering
  \includegraphics[width=8cm]{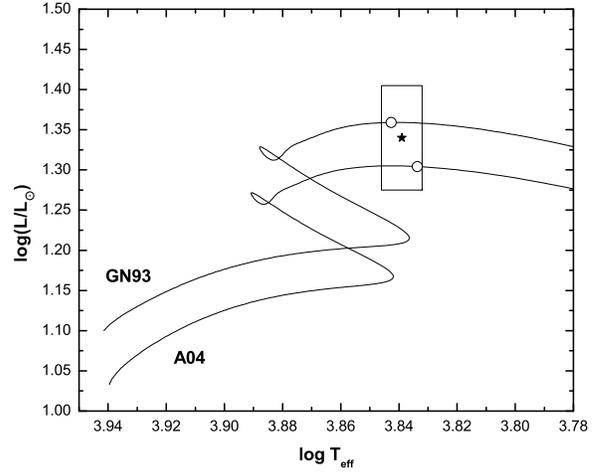}
  \caption{HR diagram showing the evolutionary tracks of the models presented in Fig.~\ref{fig:pet_mix}.}
  \label{fig:hrd_mix}
\end{figure}

Recently, Asplund et al. (\cite{asplund2004}) found evidence for a revision of the solar chemical composition (see also Asplund et al. \cite{asplund2005}). Their results indicate that the solar abundances of carbon, nitrogen, oxygen and other heavy elements are lower than previously estimated. We also tested the influence of the new solar element mixture (hereby referred to as A04) on the OPAL models of 44~Tau. 
In the Petersen diagram in Fig.~\ref{fig:pet_mix} we compare our standard post-MS model with models computed with A04 mixture and using the new solar values X=0.74, Z=0.012. As can be seen, only a small change in mass has to be applied to fit the period ratio. However, during the main sequence evolution, the element mixture itself has a significant effect on the period ratio as well. In the HR diagram (Fig.~\ref{fig:hrd_mix}) both models can be fitted within or close to the photometric error box. Furthermore, we obtain the same mode identification as for the models computed with the GN93 mixture. The instability range is only slightly shifted to lower frequencies, still predicting frequency f$_{10}$ not to be excited. It is possible to fit the observed frequencies at almost the same accuracy as for the GN93 mixture.
Additional tests with main sequence models provided similar conclusions as in the case of post-MS models.

It is important to note that the effects of both opacities and element mixture on the period ratio are comparable or even larger than rotational effects.


\section{Conclusions}
\label{sec:conclusions}

We presented a comprehensive asteroseismic study for the slowly rotating $\delta$ Scuti star 44~Tau. The spectroscopically determined $\log g$ value of $3.6 \pm 0.1$ allows no definitive determination of the evolutionary stage, so the main sequence and post-main sequence case were considered. 

Our mode degree identification based on photometric and spectroscopic data determines f$_1$ and f$_5$ as radial fundamental and first overtone modes. This confirms the results of previous studies. Four frequencies, f$_2$, f$_3$, f$_4$ and f$_6$, were identied as $\ell=1$ modes, and three frequencies, f$_7$, f$_8$ and f$_{10}$, as $\ell=2$ modes. We used the simplest model of convection (mixing-length theory) and the simplest convection-pulsation treatment (frozen convective flux assumption) which may not be adequate for cold $\delta$~Scuti stars, as discussed by Montalb\'{a}n \& Dupret (\cite{montalban}). However, our results are in good agreement with those obtained by Garrido et al (\cite{garrido}) using time-dependent convection.

The identification of the radial fundamental and first overtone frequency poses an important constraint to the models as it reduces the number of possible models significantly. Moreover, it is likely that an avoided crossing of nonradial modes in 44~Tau is observed. This fact puts additional strong constraints on the parameter space for appropriate models. For both considered evolutionary stages it is possible to fit the observed radial period ratio.

If standard chemical composition and no overshooting from the convective core are applied, only post-main sequence models of 44~Tau are located within the photometric error box in the HR diagram. Main sequence models can only be obtained by assuming convective overshooting. Examining such hydrogen-core burning models within a wide parameter range we find that all these models are significantly too cool. This fact is a major problem for main-sequence models. Nevertheless, it is remarkable that all four identified $\ell=1$ modes can be fitted in both evolutionary stages.

The predicted range of unstable modes is in good agreement with the observed frequencies in both cases. For main sequence models the instability range extends to lower frequencies than observed while for post-main sequence models higher frequencies are found to be unstable. The lowest frequency f$_{10}$ is predicted to be stable in the post-main sequence models, but it is close to the theoretical instability region. Our results on the instability ranges can be used to put a further constraint on the determination of the evolutionary stage if new frequencies are detected that extend the frequency range. Additional frequencies beyond the highest observed frequency may also help to constrain the efficiency of convection more precisely.

Considering the good fit of the observed frequencies and physical parameters such as effective temperature, we find that the post-main sequence stage is preferable to explain the pulsation spectrum of 44~Tau.

In this investigation, we also tested the effect of opacities and element mixtures on the radial modes. The choice between OPAL and OP tables causes a significant difference in the Petersen diagram. In the case of 44~Tau, the OPAL opacities are preferable for the asteroseismic analysis, because with the OP data it is not possible to fit the radial modes of 44~Tau within or close to the observational error box in the HR diagram. The use of the new solar element mixture results in a slightly cooler post-main sequence model. We do not find indications in favor of the old or the new element mixture. However, it is important to note that the influence of different opacity data on the period ratio should also be considered when using the Petersen diagram as a diagnostic tool.

\begin{acknowledgements}
We thank Wolfgang Zima for providing us with the results of the frequency analysis of the 2004 spectroscopic data. We would also like to thank Wojciech Dziembowski for valuable discussions. This work was supported by the Austrian FWF. AAP acknowledges partial financial support from the Polish MNiSW grant No. 1 P03D 021 28.
\end{acknowledgements}

\end{document}